\documentclass[prc,amsmath,showkeys,showpacs,superscriptaddress,floatfix,letterpaper]{revtex4}

\usepackage{dcolumn}
\usepackage{graphicx}

\begin {document}
  \newcommand {\nc} {\newcommand}
  \nc {\beq} {\begin{eqnarray}}
  \nc {\eeq} {\nonumber \end{eqnarray}}
  \nc {\eeqn}[1] {\label {#1} \end{eqnarray}}
  \nc {\eol} {\nonumber \\}
  \nc {\eoln}[1] {\label {#1} \\}
  \nc {\ve} [1] {\mbox{\boldmath $#1$}}
  \nc {\mrm} [1] {\mathrm{#1}}
  \nc {\half} {\mbox{$\frac{1}{2}$}}
  \nc {\thal} {\mbox{$\frac{3}{2}$}}
  \nc {\fial} {\mbox{$\frac{5}{2}$}}
  \nc {\la} {\mbox{$\langle$}}
  \nc {\ra} {\mbox{$\rangle$}}
  \nc {\etal} {\emph{et al.\ }}
  \nc {\eq} [1] {(\ref{#1})}
  \nc {\Eq} [1] {Eq.~(\ref{#1})}
  \nc {\Ref} [1] {Ref.~\cite{#1}}
  \nc {\Refc} [2] {Refs.~\cite[#1]{#2}}
  \nc {\Sec} [1] {Sec.~\ref{#1}}
  \nc {\chap} [1] {Chapter~\ref{#1}}
  \nc {\anx} [1] {Appendix~\ref{#1}}
  \nc {\tbl} [1] {Table~\ref{#1}}
  \nc {\fig} [1] {Fig.~\ref{#1}}

  \nc {\bfig} {\begin{figure}}
  \nc {\efig} {\end{figure}}
  \nc {\ex} [1] {$^{#1}$}
  \nc {\Sch} {Schr\"odinger }
  \nc {\flim} [2] {\mathop{\longrightarrow}\limits_{{#1}\rightarrow{#2}}}
\title{Influence of the projectile description
on breakup calculations}
\author{P.~Capel}
\email{pierre.capel@centraliens.net}
\affiliation{TRIUMF, 4004 Wesbrook Mall, Vancouver, B.C., Canada V6T2A3}
\author{F.~M.~Nunes}
\email{nunes@nscl.msu.edu}
\affiliation{National Superconducting Cyclotron Laboratory
and Department of Physics and Astronomy,
Michigan State University, East Lansing, Michigan 48824}
\date{\today}
\begin{abstract}
Calculations of the breakup of $^8$B and $^{11}$Be are performed
with the aim of analyzing their sensitivity to the
projectile description.
Several potentials adjusted on the same experimental data
are used for each projectile.
The results vary significantly with the potential choice,
and this sensitivity differs from one projectile to the other.
In the $^{8}$B case,
the breakup cross section is approximately scaled by 
the asymptotic normalization coefficient
of the initial bound state (ANC).
For $^{11}$Be, the overall normalization of the 
breakup cross section is no longer solely determined by the ANC.
The partial waves describing the continuum
are found to play a significant role in this variation, as
the sensitivity of the phase shifts to the projectile description
changes with the physical constraints imposed to the potential.
\end{abstract}
\maketitle

\section{Introduction}
Since the early days of radioactive beam experiments, breakup
reactions have been an important source of information 
on the structure of nuclei near the dripline.
\cite{HJJ95,Tan96,BHT03,Jon04,BR96}.
Usually, the exotic nucleus---the projectile---is simulated
by a loosely bound two-body system.
The spectroscopic factor of that single-particle
wave function is obtained from the comparison
of the theoretical predictions with the data
\cite{Nak94,Pal03,Pra03,Fuk04}.
Alternatively,
it has been suggested \cite{TCG01,TCG04} that asymptotic normalization
coefficients (ANC) can be obtained from breakup measurements.
The basic idea therein is that breakup reactions of loosely bound
nuclei are highly peripheral.
Since the asymptotic behavior
of the  projectile wave function is in general well known,
the breakup cross section is proportional to the square of the ANC.
In this work, we examine
the validity of such procedures. We focus on the dependence
of the normalization of the Coulomb induced breakup cross section on
the single particle potential used for the description of the 
ground state, as well as the dependence on other features of the 
projectile, such as excited or scattering states.

There are a variety of state-of-the-art theoretical models to describe 
a two-body projectile breaking up
in the Coulomb and nuclear fields of a
target (for a review see \cite{AN03}). It should be noted that,
although there are ways of dealing with three-body projectiles 
\cite{CAT97,Kmsu05} here we will focus on two-body projectiles only. 
Regardless of the range of applicability,
all these reaction models rely on an effective interaction between
the two bodies that constitute  the nucleus under study.
Typically, the geometry of this interaction is fixed
(a Woods-Saxon potential with $r_0 \approx 1.2$~fm 
and $a \approx 0.6$~fm) and the depth
is fitted to the binding energy of the projectile taking into
account the correct angular momentum and possibly a spin-orbit
force. It is clear that this potential is by no means unique
and the uncertainty on the potential is expected to have an impact on 
the normalization of the breakup cross section.
Within an effective-range theory, Typel and Baur have indeed
shown that the electromagnetic strengths
of halo nuclei depend on properties of the two-body
projectile description \cite{TB04,TB05}.
These include bound-state properties, like the ANC,
and parameters that characterize the final state interaction,
like the scattering lengths.
In \Ref{MN05}, it has also been found that the transfer cross section
depend significantly on the single particle parameters.
This dependence leads the authors to review the usual method
for extracting spectroscopic factors from transfer reactions.

Since the two-body system is loosely bound,
one may think that the cross section should scale with the square of
the ANC. Many questions remain open:
is the reaction sufficiently peripheral for
a dependence on ANC? To what extent
is this true when couplings in the continuum are important? 
Or, how does the final state interaction affect the results? 
In this work we will look at Coulomb breakup from several
angles: we study a proton halo $^8$B and a neutron halo $^{11}$Be,
and look at the low energy regime ($<10$~MeV/u) versus the higher energy 
regime (50--100~MeV/u). For this purpose, we use
the Continuum Discretized Coupled Channel method \cite{NT99}
or the time-dependent technique \cite{CBM03c},
according to applicability and feasibility.
Other reaction models like DWBA or
first order semiclassical theory \cite{ABH56,AW75}
are also used as qualitative tools to understand the results.
In \Sec{theory}, we present a brief description of the reaction
models to be used. In \Sec{potentials}, the interactions describing 
realistic cases are specified
as well as other details concerning the calculations.
\Sec{bu} contains the results of the calculations of the Coulomb
breakup of $^8$B and $^{11}$Be. In \Sec{analysis},
tests on the sensitivity to the single particle parameters of the projectile
are presented and  discussed.
Finally in \Sec{conclusion}, conclusions are drawn.

\section{Theoretical framework}\label{theory}
\subsection{Projectile description}\label{H0}

Within the theoretical description of
the breakup of a halo nucleus $P$ impinging on a target $T$,
the projectile is usually assumed to have a two-body structure:
a pointlike and structureless fragment $f$
(of mass $m_f$ and charge $Z_f e$) loosely bound to a
structureless core $c$ (of mass $m_c$ and charge $Z_c e$).
The target is described as a structureless particle of mass
$m_T$ and charge $Z_Te$.
The internal structure of the two-body projectile is
described by the Hamiltonian $H_0$
\beq
H_0=-\frac{\hbar^2}{2\mu}\Delta+V_{cf}(\ve{r}),
\eeqn{e1}
where $\ve{r}$ is the relative coordinate of the fragment to
the core, and $\mu=m_cm_f/(m_c+m_f)$ is the reduced mass
of the $c$-$f$ system.
The potential $V_{cf}$ simulates the interaction between
the core and the fragment. It is composed of a Coulomb term
plus a nuclear term, which comprises a central part and
a spin-orbit coupling term
\beq
V_{cf}(\ve{r})=V_C(r,R_C)+V_0(r)+\ve{l}\cdot\ve{I}V_{lI}(r),
\eeqn{e2}
where $\ve{l}$ is the orbital momentum of the $c$-$f$
relative motion, and $\ve{I}$ is the spin of the fragment.
The spin of the core is neglected and assumed to be nil.

The Coulomb term $V_C$ corresponds to the potential due
to a uniformly charged sphere of radius $R_C$ (the core)
acting on a pointlike particle (the fragment)
\beq
V_C(r,R_C)=\left\{
\begin{array}{ll}
\frac{1}{2}\frac{Z_cZ_f e^2}{R_C}\left(3-\frac{r^2}{R_C^2}\right)&r<R_C\\
\frac{Z_cZ_f e^2}{r}&r\ge R_C.
\end{array}
\right.
\eeqn{e2a}
The central part of the nuclear potential $V_0$ has a Woods-Saxon form factor
\beq
V_0(r)=-V_l f(r,R_0,a),
\eeqn{e3}
where
\beq
f(r,R_0,a)=\left[1+\exp\left(\frac{r-R_0}{a}\right)\right]^{-1}.
\eeqn{e4}
The spin-orbit coupling term has the usual Thomas form factor
\beq
V_{lI}(r)=V_{lS}\frac{1}{r}\frac{d}{dr}f(r,R_0,a).
\eeqn{e5}
The radius is parameterized as: $R_0=r_0A_c^{1/3}$,
where $A_c$ is the mass number of the core.
The depths of the potential are adjusted to reproduce the
bound states of the system and some of its resonances.

In the $lj$ partial wave, the eigenstates of $H_0$ are
\beq
H_0\phi_{ljm}(E,\ve{r})=E \phi_{ljm}(E,\ve{r}),
\eeqn{e6}
where $j$ is the total angular momentum resulting from
the coupling of the orbital momentum $l$ and the
fragment spin $I$, and $m$ is its projection.
The negative-energy states correspond either to physical
bound states or to Pauli-forbidden states simulating
the presence of fragment-like particles in the core.
They are normed to unity.
The positive energy states describe the continuum of the
two-body projectile, i.e. the scattering of the fragment by the core.
Their radial part $r^{-1} u_{lj}$ is normalized as
\beq
u_{lj}(k,r)\flim{r}{\infty}
[\cos\delta_{lj}(k)F_l(k,r)
+\sin\delta_{lj}(k)G_l(k,r)],
\eeqn{e7}
where $k=\sqrt{2\mu E/\hbar^2}$ is the wave number of the
relative motion of the core and the fragment,
$F_l$ and $G_l$ are respectively the regular and irregular
Coulomb functions \cite{AS70}, and $\delta_{lj}$ is the nuclear phase-shift.

\subsection{CDCC}\label{CDCC}

The reaction of a loosely bound two-body projectile on a target
can be approximated to a three-body scattering problem. Typically,
target excitation is neglected and
the three-body wave function is expanded in terms of the intrinsic
motion of the core and the fragment $c+f$ 
within the projectile $u_{lj}(k,r)$,
and the motion of the projectile relative to the target
$f_{LjJ}(k,R)$. Here $\ve{R}$ is the vector connecting the center of mass
of the projectile with the target,
$L$ is  the corresponding orbital angular momentum,
and $J$ is the total angular momentum of the system. 

For modeling 
breakup it is important to have a good description of the projectile
continuum. One of the most successful methods for elastic breakup
is the Continuum Discretized Coupled Channel method 
(CDCC) \cite{SYK86} first developed for deuteron breakup but
today widely applied to exotic nuclei
(e.g. \cite{NT99,TNT01,Tos02,TTS03,SN04,Oga05}).
In the standard CDCC, the projectile continuum is included through 
a discretization into energy bins. For the bin wave function,
we use the integral over the momentum interval, weighted by the 
function $g_ {lj}$
in the following way :
\begin{equation} \label{binwf}
{  \widetilde{u}_{lj,i} (r) = \sqrt {\frac{2 }{\pi N_{lj}}} ~~
        \int_{k_i}^{k_{i+1}} g_{lj}(k) u _{lj} (k,r) dk}\;.
\end{equation}
For further details on the normalization $N_{lj}$ of the bins 
see \Ref{TNT01}.

The three-body Schr\"odinger equation can then be reduced
to a coupled channel equation in $R$,
\begin{eqnarray} \label{crc}
 \left [ - \frac{\hbar^2 }{2 \mu} ~ \left ( \frac{d^2 }{ dR^2} - 
 \frac{L(L+1)} {R^2} \right ) + V^J _{\alpha:\alpha}(R) 
 +E_i - \mathcal{E} \right ] f _ {\alpha J} (R) 
= \sum _ {\alpha ' \ne \alpha} 
 i ^ {L ' - L} ~ V^J _{\alpha:\alpha'}(R)  f_{\alpha' J} (R) ,
\end{eqnarray}
where the subscript $\alpha$ represents all relevant quantum numbers
and $E_i$ is the average energy of continuum bin $[k_i,k_{i+1}]$
(or  $E<0$ for the bound states),
and $\mathcal{E}$ is the total energy of the system.
The coupling potential $V^J _{\alpha:\alpha'}(R)$
consists of the sum of the core-target and fragment-target interactions
averaged over the projectile states \cite{NMR04}.
One can solve \Eq{crc} exactly: the solution $\Psi^{CDCC}$
includes all possible rearrangements within the projectile continuum.
One can also solve \Eq{crc} iteratively: the one-step 
iteration of \Eq{crc} is equivalent  to the standard
DWBA, where the optical potential between the projectile and
the target is replaced by the folding potential $V^J _{0:\alpha'}(R)$
\cite{NT98}.

\subsection{Time-dependent model}\label{TDSE}

The time-dependent description of the reaction
\cite{KYS94,EBB95,EB96,MB99,TW99,Fal02,CBM03c} relies
on the semiclassical approximation \cite{ABH56,AW75} in which the
projectile-target relative motion is treated classically,
while the internal motion of the projectile is described
quantum mechanically.
The target is assumed to follow a classical trajectory
in the projectile rest frame.
The projectile is therefore seen
as evolving in a time-dependent potential which simulates
its interaction with the target. The wave function $\Psi$
describing its internal structure is
solution of the following time-dependent \Sch equation
\beq
i\hbar\frac{\partial}{\partial t}\Psi(\ve{r},t)
=\left[H_0+V(\ve{r},t)\right]\Psi(\ve{r},t),
\eeqn{e10}
where $H_0$ [see \Eq{e1}] describes the internal structure of the projectile,
and $V$ is the time-dependent potential that simulates the $P$-$T$ interaction.
The latter reads
\beq
V(\ve{r},t)=V_{cT}[r_{cT}(t)]+V_{fT}[r_{fT}(t)]-\frac{(Z_c+Z_f)Z_Te^2}{R(t)},
\eeqn{e11}
where the time-dependent coordinate $\ve{R}$ describes
the classical trajectory followed by the target in the
projectile rest frame.
It is a hyperbola, which can be approximated fairly well by
a straight line at high energies.
The vectors $\ve{r}_{cT}$ and $\ve{r}_{fT}$ correspond respectively
to the core-target and fragment-target coordinates.
In \Eq{e11}, $V_{cT}$ and $V_{fT}$ are local potentials
which model the interaction between the target and the
projectile constituents. They comprise a Coulomb term
and a short-range optical potential, which
simulates the nuclear interaction.

\Eq{e10} is solved numerically considering the projectile
initially in its ground state $\phi_{l_0j_0m_0}$ of energy $E_0$.
We use the algorithm described in \Ref{CBM03c}, and consider
linear trajectories.
For each trajectory, characterized by impact parameter $b$,
we deduce the breakup probability $P_{\mrm{bu}}$ by projecting the output
wave function $\Psi(\ve{r},t\rightarrow+\infty)$ onto the positive
eigenstates of $H_0$, which describe the projectile after dissociation.
The breakup cross section $\sigma_{\mrm{bu}}$ is then obtained by summing
this probability over all impact parameters.

At sufficiently high energies, the solution of
\Eq{e10} can be approximated by
first-order perturbation theory \cite{ABH56,AW75}.
In that approximation the breakup process is assumed to occur
in one step from the initial ground state to the continuum,
and the couplings inside the continuum are neglected.
The first-order approximation of the breakup cross section reads
\beq
\frac{d\sigma^{(1)}_{\mrm{bu}}}{dE}(E)=\frac{1}{\hbar^2}\frac{2\pi}{2j_0+1}
\sum_{m_0}\sum_{ljm}\int \left|\left\langle\phi_{ljm}(E,\ve{r})\left|
\int_{-\infty}^{\infty}e^{i\omega t}V(\ve{r},t)dt
\right|\phi_{l_0j_0m_0}(E_0,\ve{r})\right\rangle\right|^2 d\ve{b},
\eeqn{e21}
where $\omega=(E-E_0)/\hbar$.

When the projectile-target interactions are purely
Coulomb,
the first-order approximation of the breakup probability
can be easily expressed as a series of the contributions
of each multipole of the time-dependent potential \eq{e10}.
In practice, only the first terms of that series are needed.
For straight-line trajectories, a semi-analytical expression
of the dipole and quadrupole
contributions to the first-order
breakup probability can be obtained (see e.g. \Ref{CB05}).

\section{Two-body interactions}\label{potentials}

\subsection{Description of \ex{8}B}\label{B8}

The \ex{8}B nucleus is described by the usual two-body system:
a proton loosely bound to a \ex{7}Be core in its $\thal^-$
ground state \cite{TB94,EB96,TWB97,TNT01,MTT02}.
The internal structure of the core is neglected,
and its spin is set to zero in the calculations.
The $2^+$ ground state of \ex{8}B is assumed to be
a pure $p3/2$ proton single-particle state.
As explained in \Sec{H0}, the interaction between
the core and the proton is simulated by a local
potential [see Eqs.~\eq{e2}--\eq{e5}].

In order to study the sensitivity of the calculations
to the description of the projectile,
five \ex{7}Be-$p$ potentials are considered.
The parameters of these potentials are listed in \tbl{t01}.
They are obtained by varying the diffuseness
of a simplified version of the potential
developed by Esbensen and Bertsch \cite{EB96} (potential T2).
While the spin-orbit term is kept unchanged for all
potentials, the depth of the central part $V_l$ is
adjusted to reproduce the 137~keV binding energy of the
bound state \cite{Ajz88}. 
Since the spin of the core is neglected, the $0^+$,
$1^+$, and $3^+$ states resulting from the coupling of
this $\thal$ spin and the $\frac{3}{2}$ angular momentum of the
halo proton are degenerate with the $2^+$ ground state.
Therefore, the $p$-wave $1^+$ resonance at $E=0.63$~MeV \cite{Ajz88}
is not reproduced by this parameterization.
We use the same potential for all partial waves.
\begin{table}
\begin{tabular}{l c c c c}
{Potential} &
{$V_l$} & {$V_{lS}$} & {$a$} & {$r_0$}\\
 & (MeV) & (MeV fm\ex{2}) & (fm) & (fm)\\ \hline
T1 & 45.23  & 19.59 & 0.40 & 1.25\\
T2 & 44.98  & 19.59 & 0.52 & 1.25\\
T3 & 44.47  & 19.59 & 0.60 & 1.25\\
T4 & 43.50  & 19.59 & 0.70 & 1.25\\
T5 & 42.28  & 19.59 & 0.80 & 1.25\\
\end{tabular}
\caption{Parameters of the $^7$Be-$p$ potentials
[see Eqs. \protect\eq{e2}-\protect\eq{e5}].
Note that $R_0$ used in \protect\eq{e2a}-\protect\eq{e5}
is parameterized as $r_0 A_c^{1/3}$, and the Coulomb radius is always
kept constant at $R_C=1.3 A_c^{1/3}$ fm.}
\label{t01}
\end{table}
The levels obtained with the potentials of \tbl{t01}
are listed in \tbl{t02}. The ANC of the ground state
$b_{0p3/2}$ is given as well.
Note that besides the $0p3/2$ ground state, the potentials
exhibit also a $p1/2$ resonance, which
does not correspond to any known physical state.
This unphysical state has not been adjusted,
and its energy varies from one potential to the other.

\begin{table}
\begin{tabular}{l | c c | c c}
Potential&$E_{0p3/2}$&$b_{0p3/2}$&$E_{0p1/2}$&$\Gamma_{0p1/2}$\\
         & (MeV) &   (fm$^{-1/2}$)& (MeV)     &    (MeV)        \\
 \hline
T1 &-0.137 & 0.6477 & 2.8 & 2.2\\
T2 &-0.137 & 0.7008 & 2.3 & 1.6\\
T3 &-0.137 & 0.7410 & 1.9 & 1.2\\
T4 &-0.137 & 0.7959 & 1.6 & 0.9\\
T5 &-0.137 & 0.8554 & 1.3 & 0.7\\
\end{tabular}
\caption{Levels obtained with the five potentials listed in
\protect\tbl{t01}. The energies and widths are expressed in MeV.
The ANC of the $0p3/2$ ground state $b_{0p3/2}$ is displayed as well.}\label{t02}
\end{table}

\subsection{Description of \ex{11}Be}\label{Be11}

As in previous studies \cite{KYS94,EBB95,MB99,TS01r,Fal02,CBM03c,Zad04,CGB04},
\ex{11}Be is described as a neutron loosely bound
to a \ex{10}Be core. The \ex{10}Be core is assumed to be in its
$0^+$ ground state.
The depths of the $V_{cf}$ potential \eq{e2}
simulating the interaction between \ex{10}Be and the neutron
are adjusted to reproduce the low-lying levels in the \ex{11}Be
spectrum \cite{Ajz90}. The well known shell inversion observed between
the two bound states of \ex{11}Be is reproduced using
a parity dependent depth of the central term of the nuclear
potential $V_l$.
The $\half^+$ ground state is modeled by a $1s1/2$ state, the
$\half^-$ excited state by a $0p1/2$ state, and the $\fial^+$
resonance is adjusted in the $d5/2$ partial wave.

In this case, we make use of six sets of parameters
to study the sensitivity of our calculations to the
potential choice.
They are listed in \tbl{t1}.
The first potential (V1) was developed
for a time-dependent analysis of the breakup of \ex{11}Be
on \ex{12}C \cite{CGB04}.
The next four (V2 to V5) have been derived from V1 by varying either
the diffuseness or the radius of the Woods-Saxon form factor \cite{Cmsu05}.
The values were chosen to encompass those used in most other
breakup calculations \cite{KYS94,EBB95,TS01r,Fal02}.
In addition, we also use a sixth potential (V6)
developed by Fukuda \etal for analyzing their data \cite{Fuk04}.
It reproduces only the ground state energy of \ex{11}Be and
does not include a spin-orbit coupling term.

\begin{table}
\begin{tabular}{c c c c c c}
{Potential} &
{$V_{l \mrm{even}}$} &
{$V_{l \mrm{odd}}$} &
{$V_{lS}$} &
{$a$} &
{$r_0$}\\
 & (MeV) & (MeV) & (MeV fm\ex{2}) & (fm) & (fm)\\ \hline
V1 & 62.52 & 39.74 & 21.0 & 0.6 & 1.2\\
V2 & 66.325 & 38.37 & 12.44 & 0.5 & 1.2\\
V3 & 58.905 & 40.025 & 27.68 & 0.7 & 1.2\\
V4 & 71.28 & 49.015 & 29.95 & 0.6 & 1.1\\
V5 & 55.25 & 32.515 & 12.86 & 0.6 & 1.3\\
V6 & 59.05 & 59.05 & 0 & 0.62 & 1.236
\end{tabular}
\caption{Parameters of the $^{10}$Be-$n$ potentials
[see Eqs. \protect\eq{e2}-\protect\eq{e5}].
Note that $R_0$ used in \protect\eq{e2a}-\protect\eq{e5}
is parameterized as $r_0 A_c^{1/3}$.}
\label{t1}
\end{table}

The physical energy levels
obtained with these six potentials are listed
in \tbl{t2}.
The asymptotic normalization coefficient $b_{1s1/2}$
of the ground state is also given.

\begin{table}
\begin{tabular}{c| c c| c| c c}
Potential&$E_{1s1/2}$&$b_{1s1/2}$&$E_{0p1/2}$&$E_{0d5/2}$&
$\Gamma_{0d5/2}$\\ 
 & (MeV) & (fm\ex{-1/2}) & (MeV) & (MeV) & (MeV) \\ \hline
V1&-0.504&0.83&-0.184&1.274&0.162\\
V2&-0.504&0.80&-0.184&1.274&0.131\\
V3&-0.504&0.87&-0.184&1.274&0.200\\
V4&-0.504&0.82&-0.184&1.274&0.145\\
V5&-0.504&0.85&-0.184&1.274&0.181\\
V6&-0.504&0.85&-12.5 & 2.6 &0.9  

\end{tabular}
\caption{Levels obtained with the six potentials listed in
\protect\tbl{t1}. Potentials V1 to V5
were adjusted to the first three levels of \ex{11}Be,
while V6 reproduces only the $\half^+$
ground state. The ANC of the $1s1/2$ state
is displayed as well.}\label{t2}
\end{table}

\subsection{Projectile-target interactions}
The reactions  we consider are Coulomb dominated.
However, the nuclear interactions between the target
and the projectile composites are not negligible \cite{TNT01,CBM03c}.
As mentioned earlier, they are simulated by
short ranged optical potentials.
The general form factor of the potential simulating
the interaction between the constituent $x$ and the target $T$ is
\beq
V_{xT}(r)=V_C(r,R_C)-Vf(r,R_R,a_R)
-iWf(r,R_I,a_I)-iW_D a_I\frac{d}{dr}f(r,R_I,a_I),
\eeqn{e30}
where $V_C$ is the point-sphere Coulomb potential \eq{e2a},
and $f$ is the Woods-Saxon form factor \eq{e4}.
The values of the parameters we consider for these interactions
are listed in \tbl{t3}. They correspond to potentials
found in the literature for similar systems.

As in \Ref{TNT01}, we simulate the interaction between
\ex{7}Be and \ex{58}Ni at low energy
using the potential developed by Moroz \etal \cite{Mor82}.
This potential reproduces elastic scattering data of
\ex{7}Li impinging on \ex{58}Ni at 14.2~MeV.

To model the interaction between \ex{10}Be and \ex{208}Pb
at 69~MeV per nucleon, as in \Ref{CBM03c},
we follow Typel and Shyam \cite{TS01r}
and adapt a potential developed by Bonin \etal \cite{Bon85},
which reproduces the elastic scattering cross section
of $\alpha$ on lead at 699 MeV.

The interactions of the valence nucleon with the targets are
approximated by the parameterization of Becchetti and Greenlees
\cite{BG69}, as in previous works \cite{TNT01,CBM03c}.

\begin{table}
\begin{tabular}{c c c|c c c c c c c c}
   &   &Energy       &$V$  &$R_R$&$a_R$&$W$  &$W_D$   &$R_I$&$a_I$&$R_C$\\
$x$&$T$&(MeV/nucleon)&(MeV)&(fm) &(fm) &(MeV)&(MeV)&(fm) &(fm) &(fm)\\
\hline
\ex{7}Be &\ex{58}Ni &3&100   &4.064&0.65&30.6& 0    &4.347&0.80&5.032\\
\ex{10}Be&\ex{208}Pb& 69&70.0  & 7.43&1.0 &58.9& 0    &7.19 &1.04&5.92 \\
$p$      &\ex{58}Ni &3&54.512&4.529&0.75&0   &11.836&4.877&0.58&5.032\\
$n$      &\ex{208}Pb& 69&29.46 & 6.93&0.75&13.4& 0    &7.47 &0.58& - \\
\end{tabular}
\caption{Parameters of the optical potentials \protect\eq{e30} which
simulate the interaction between the projectile constituents and the
targets.}\label{t3}
\end{table}

\section{Breakup calculations}\label{bu}

\subsection{The $^8$B case}\label{b8bu}
The breakup of \ex{8}B on \ex{58}Ni has been measured
at 25.75~MeV at Notre-Dame \cite{Gui00,Kol01}.
This reaction is analyzed in \Ref{TNT01}
within the CDCC framework (see \Sec{CDCC}).
The calculation is in good agreement with the data using
the two-body description of \ex{8}B given in \Sec{B8}
(potential T2 of \tbl{t01}).

In the present paper, we study the sensitivity
of this analysis to the \ex{7}Be-$p$ potential choice.
CDCC calculations identical to that of \Ref{NT99} are performed
with the code FRESCO \cite{fresco}
considering the five potentials T1--T5 listed in \tbl{t01}.
For the model space, we use $L$ up to 1000, a maximum radius
for the distorted waves of $R_{\mrm{max}}=500$ fm and coupling
potentials truncated at $R_{\mrm{bin}}=60$ fm. The discretization
of the continuum of the projectile is made up to $E_{\mrm{max}}=4$ MeV
and all dipole and quadrupole transitions for s, p and d waves are included.
As in \Ref{NT99}, the nuclear interactions between the projectile constituents
and the nickel target are simulated by the optical potentials
given in \tbl{t3}.

First, let us consider simpler cases of the continuum wave functions
whilst changing the ground state interaction according to models T1--T5.
If only pure Coulomb waves are taken in the \ex{7}Be-$p$ continuum,
the scaling with the square of the ANC of the initial bound state
$b_{0p3/2}$ is perfect (i.e. with less than 1~\% difference).
This same result is obtained
if the  nuclear scattering  \ex{7}Be-$p$ potential is kept fixed to T1
for all models.
We have checked these results using other reaction models.
The DWBA approximation leads to the same conclusion,
as well as the first-order approximation when the breakup of \ex{8}B on
\ex{58}Ni is computed at higher energy (50~MeV/nucleon)
considering a purely Coulomb interaction between the projectile and the target.
The direct proportionality of the cross section to the square of the ANC
of the initial bound state indicates that only the asymptotic part of the
wave function has an influence upon the cross section, 
consistent with the small binding energy of the projectile.

\begin{figure}
\includegraphics[height=7cm]{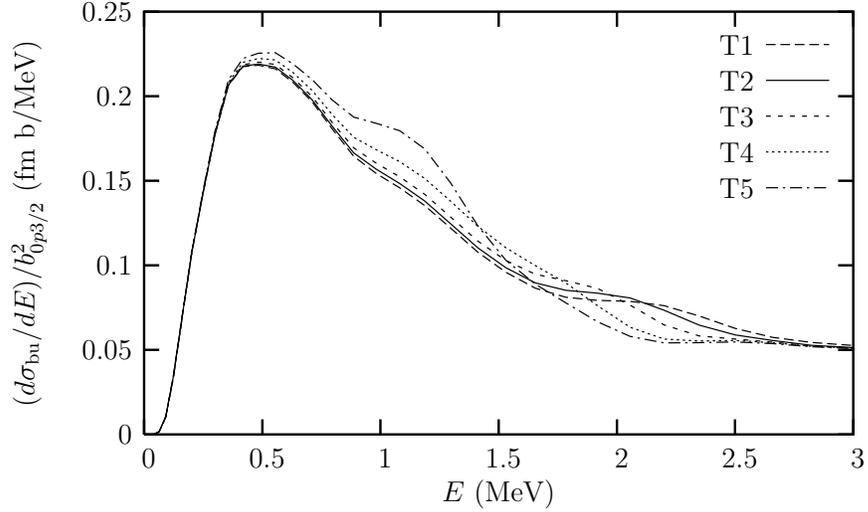}
\caption{Cross section for the breakup
of \ex{8}B on \ex{58}Ni at 25.75~MeV divided by the square of the
single particle ANC.
The cross section is given as a function of the \ex{7}Be-$p$ relative
energy $E$ after breakup.
Calculations are performed with the different \ex{7}Be-$p$
potentials listed in \tbl{t01}.}\label{f1}
\end{figure}

When the scattering potential
is chosen to correspond to the ground-state interaction,
the breakup cross section is 
only approximately proportional to $b_{0p3/2}^2$.
This is illustrated in \fig{f1}, where the breakup
cross sections obtained with potentials T1--T5 are
plotted as a function of the relative energy between the
\ex{7}Be core and the proton after breakup.
To emphasize the sensitivity to the potential choice,
those values are divided by the square of the ANC.
The different curves exhibit similar behaviors.
They all increase sharply to reach a maximum around 0.5~MeV,
and display a bump in the slow decrease that follows the maximum.
However, these behaviors are not identical.
First, even though they have been divided by $b_{0p3/2}^2$,
their magnitudes differ by approximately 4~\% in the vicinity
of the maximum.
Second, the location of the bump changes with the potential choice.
Since these features do not appear when the same scattering potential
is used, they are due to differences in the distorted waves.

\begin{figure}
\includegraphics[height=7cm]{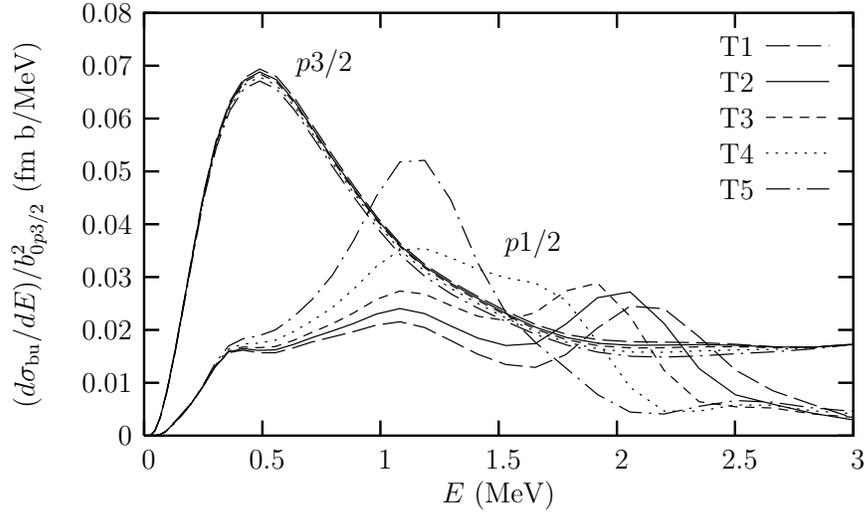}
\caption{Contributions of
$p3/2$ and $p1/2$ partial waves to the breakup cross section
of \ex{8}B on \ex{58}Ni.
They are plotted as a function of $E$, and are divided by the square of the
single particle ANC.
Calculations are performed with the different \ex{7}Be-$p$
potentials listed in \tbl{t01}.}\label{f2}
\end{figure}

To understand these features, we analyze the contributions of
the different partial waves to the breakup cross section.
The sensitivity of these contributions to the potential choice
vary from one partial wave to the other.
This is illustrated in \fig{f2}, which displays
the $p3/2$ (dominant) and the $p1/2$ contributions to the
cross section. They both have been divided by $b_{0p3/2}^2$.
On the one hand,
the $p3/2$ contributions---as well as the $s$ and $d$
ones---exhibit very similar shapes, but
their magnitudes vary by 4~\%---this variation goes
up to 8~\% for the $s$ component.
These results explain the non exact proportionality of the total 
breakup cross section
to the square of the single particle ANC.
On the other hand, each $p1/2$ contribution exhibits a maximum
whose location varies with the potential choice.
These maxima are responsible for the presence of the
bumps in the total cross sections above 1~MeV.
They are related to the $p1/2$
resonance obtained with potentials T1--T5 (see \tbl{t02}).
This effect has already been observed in the breakup of
\ex{11}Be on \ex{12}C, where the presence of the $\fial^+$
resonance in the \ex{11}Be spectrum
induces a narrow peak in the cross section \cite{CGB04,Fuk04}.
It confirms that the description of the continuum has a significant
influence on the breakup calculation.

The same effects are observed in DWBA calculations, 
and, at higher energy (50 MeV/nucleon),
within the first-order perturbation theory.
However, since the contribution of the E2 component
decreases at high energy, the sensitivity to the $p$-wave
continuum is less noticeable in the latter case.

These results indicate that the breakup calculations are
sensitive, not only to the tail of the ground-state wave function,
but also to the way the continuum of the projectile is described.
This influence of the continuum description upon breakup
calculations has already been mentioned in \Ref{TCG04}.
In that reference, Trache \etal observed that the momentum
distributions computed using distorted waves differed
from those obtained previously with plane waves \cite{TCG01}.
Since the reaction is very peripheral, the sensitivity
to the scattering potential is most likely due to the
subsequent variations in the phase shifts.
The detailed analysis of these effects is addressed in \Sec{analysis}.


\subsection{The $^{11}$Be case}\label{be11bu}
A more severe case of the breakdown of the ANC scaling
is found in the Coulomb breakup of \ex{11}Be on lead.
This reaction has recently been measured at RIKEN
at 69~MeV/nucleon \cite{Fuk04}.
From the analysis of their experiment, Fukuda \etal
extracted a spectroscopic factor of about 0.7 for the
$|^{10}\mrm{Be}(0^+)\otimes s1/2\rangle$ configuration
of the ground state of \ex{11}Be.

With the aim of analyzing the sensitivity of this
figure to the \ex{10}Be-$n$ potential, we perform
time-dependent calculations of the reactions using
the six potentials V1--V6 described in \Sec{Be11}.
As in \Ref{CBM03c}, the calculations are done within the time-dependent
framework described in \Sec{TDSE} using the optical
potentials listed in \tbl{t3} for simulating the nuclear
interactions between the projectile and the target.

\begin{figure}
\includegraphics[height=7cm]{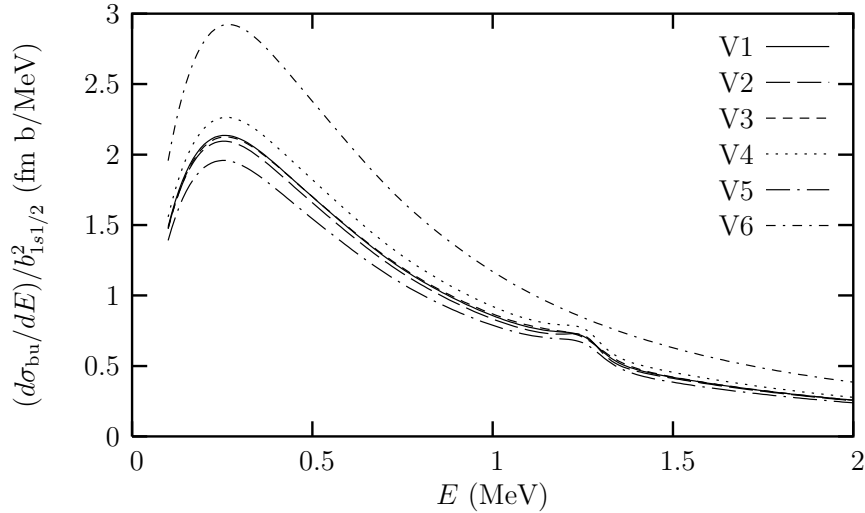}
\caption{Breakup cross section of \ex{11}Be on \ex{208}Pb
at 69~MeV/nucleon as a function of the energy $E$.
The value is divided by the square of the single particle
ANC.
Calculations are performed with potentials V1--V6 listed
in \protect \tbl{t3}.}\label{f3}
\end{figure}

As for \ex{8}B, if simplifications are performed in the
continuum (namely, switching off the nuclear Be-$n$ interaction,
or taking the same interaction for all scattering partial waves 
as the ground state), the resulting breakup cross section
is directly proportional to the square of the ground state ANC
$b_{1s1/2}^2$.
When the realistic $^{11}$Be interactions of table II are used,
the situation changes: the cross section is no longer
proportional to $b_{1s1/2}^2$.
This is illustrated in \fig{f3}, where the breakup
cross sections divided by $b_{1s1/2}^2$ are plotted
as a function of the relative energy
between the \ex{10}Be core and the neutron after breakup.
All curves exhibit the same shape---if one excepts the small
bump around 1.3~MeV due to the $d5/2$ resonance, which is not
reproduced by the potential V6.
However, their magnitude varies significantly from one potential
to the other.
In particular, V6 leads to a cross section larger by 40~\%
than V1--V5.
Moreover, even though they have been adjusted on the same
energy levels, these five potentials lead to variations in the breakup cross
sections as large as 20~\%.
Since this discrepancy is observed only when the scattering
potential differs, we conclude
that the way the continuum is described has a significant
influence on the breakup calculation.
The same result is obtained within the first-order approximation.
With such a variation of the calculation
with the potential choice, the validity of the spectroscopic
factor extracted by Fukuda \etal from their breakup measurement
\cite{Fuk04} is questionable.

\begin{figure}
\includegraphics[height=7cm]{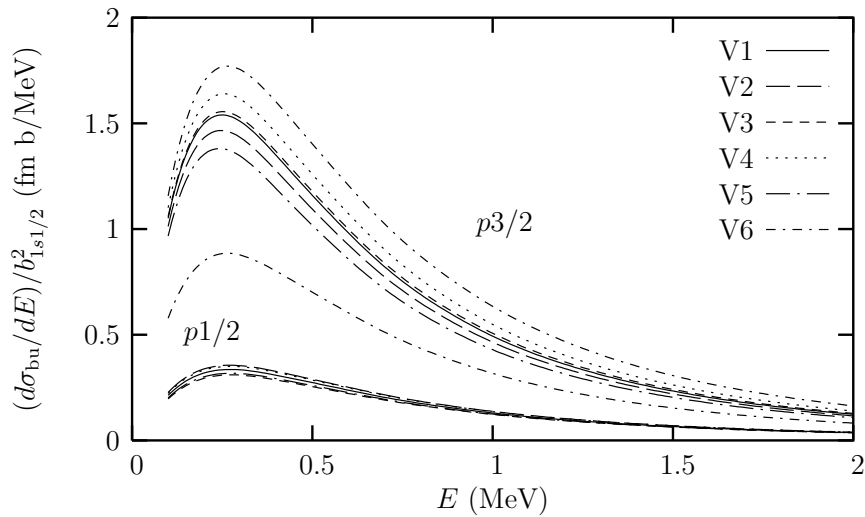}
\caption{Contributions of the main partial waves
$p3/2$ and $p1/2$ to the breakup cross section of \ex{11}Be on \ex{208}Pb.
The values are divided by the square of the single particle.
}\label{f4}
\end{figure}

The effect observed in \fig{f3} is very different from
one partial wave to the other.
This is illustrated in \fig{f4}, where the dominant $p3/2$ and $p1/2$
contributions to the total cross section are depicted.
Like the results shown in \fig{f3}, they have been divided
by $b^2_{1s1/2}$ to cancel the dependence on the ANC.
On the one hand, the difference between V1--V5 is
located mainly in the $p3/2$ contribution.
These potentials lead indeed to very similar $p1/2$ contributions.
On the other hand, even if the $p3/2$ contribution of potential V6
is still higher than that of the others,
the major difference between that potential and the other five
lies mainly in the $p1/2$ contribution.
This partial wave analysis confirms that the breakup cross
section depends not only on the ground-state wave function,
but also on the description of the partial waves describing the
continuum.

\section{Analysis of the sensitivity to the potential choice}\label{analysis}
The features presented in the previous section
are all reproduced using fully coupled reaction models
(like CDCC or time-dependent approaches) and also one-step
approximations (like DWBA or first-order perturbation theory).
This indicates that those dependences appear mainly through
first-order transitions from the initial bound state to
the continuum.
Therefore, higher-order effects will be neglected in the qualitative analysis
presented in this section.
The effects observed in section \ref{bu} can  be explained
through the study of the wave functions of the initial bound
states and those of the partial waves describing the continuum [see \Eq{e21}].
Since for both nuclei the breakup reaction is Coulomb dominated,
we neglect the nuclear contribution in the following qualitative analysis
and consider only the Coulomb potential, which we expand into multipoles.

\begin{figure}
\includegraphics[width=10cm]{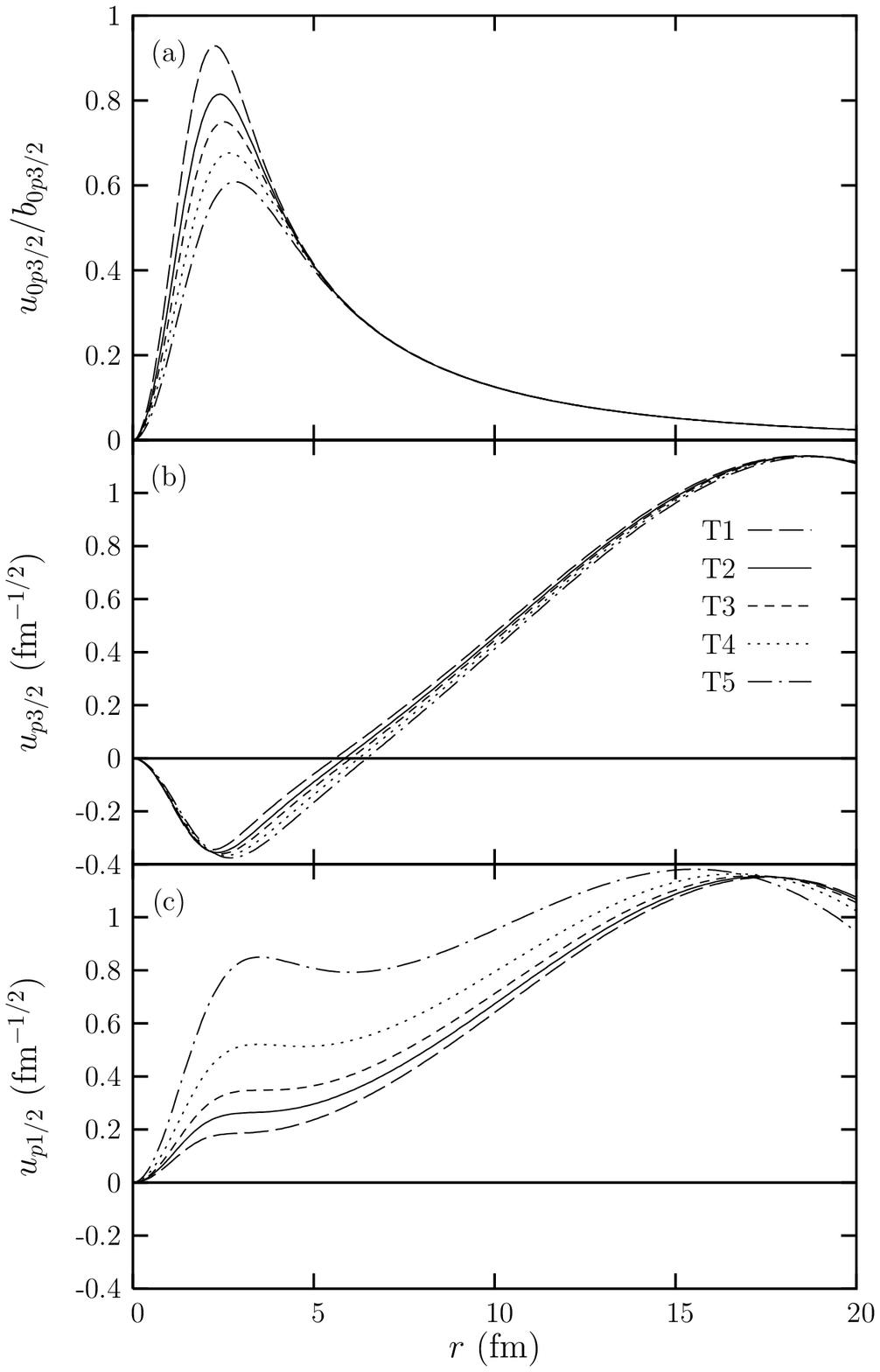}
\caption{Wave functions of \ex{8}B obtained with the different
\ex{7}Be-$p$ potentials T1--T5:
(a) the initial ground state divided by $b_{0p3/2}$,
(b) and (c) the $p3/2$ and $p1/2$, respectively, continuum wave functions
computed at $E=1$~MeV.
}\label{f5}
\end{figure}

The contributions of the $p$ waves to the breakup of
\ex{8}B on \ex{58}Ni displayed in \fig{f2} correspond
mainly to E2 transitions from the ground state.
These transitions are large because the reaction occurs at low energy.
\fig{f5} pictures the radial wave functions of the ground state (a) and
the continuum states, $p3/2$ (b) and $p1/2$ (c),
computed at an energy of 1~MeV.

Once divided by their ANC, the wave functions of the
\ex{8}B ground state obtained with the different \ex{7}Be-$p$
potentials are identical above 5~fm [\fig{f5}(a)].
The only difference between them lies below that radius.
Due to the low binding energy of the system,
these wave functions exhibit a very long range.
This long range, and the presence of the $r^2$ factor
of the quadrupole term of the Coulomb potential
in the transition matrix element ensures the peripherality
of the breakup reaction.
Consequently, the breakup cross sections are 
not very sensitive to the differences at small radius.
This explains why, when the same description of the continuum
is chosen in all calculations, the cross section is
exactly proportional to $b^2_{0p3/2}$.
If the continuum is described using
potentials T1--T5, the corresponding
wave functions differ from one potential to the other.
Since the reaction is very peripheral,
only the differences in the distorted waves
for $r \ge 5$ fm (i.e. mainly phase shifts)
lead to variations in the breakup cross sections.
This difference is illustrated in  part (b)
of \fig{f5} for the dominant $p3/2$ partial wave.
Potentials T1--T5 indeed lead to variations
in the phase shift of the wave function.
These remain small though
due to the fact that
all potentials have been adjusted to reproduce the
low-lying ground state in that partial wave.
However, they are not negligible,
and explain the loss of exact proportionality of the $p3/2$
contribution to the breakup cross section to the square of the ANC,
illustrated in \fig{f2}.
A similar analysis of the $s$ and $d$ continuum wave functions
explains the lack of proportionality to $b_{0p3/2}^2$
of these components as well.

The differences between the $p1/2$ partial waves
are much more significant [\fig{f5}(c)].
This is due to the presence in that partial wave of
a low-lying resonance whose energy varies from one potential
to the other (see \tbl{t02}).
These large variations account for the significant distortions
shown in \fig{f2}.

In the case of the Coulomb breakup of \ex{11}Be
at 69~MeV/nucleon the same reasoning can be done
in order to explain the significant variations observed in \fig{f3}.
Occurring at higher energy, the transition
to the continuum is dominated by the E1 term of
the Coulomb interaction \cite{CB05}. Therefore,
starting from an initial $s$ state, the transfer
to the continuum occurs mainly through $p$ waves.

\begin{figure}
\includegraphics[width=10 cm]{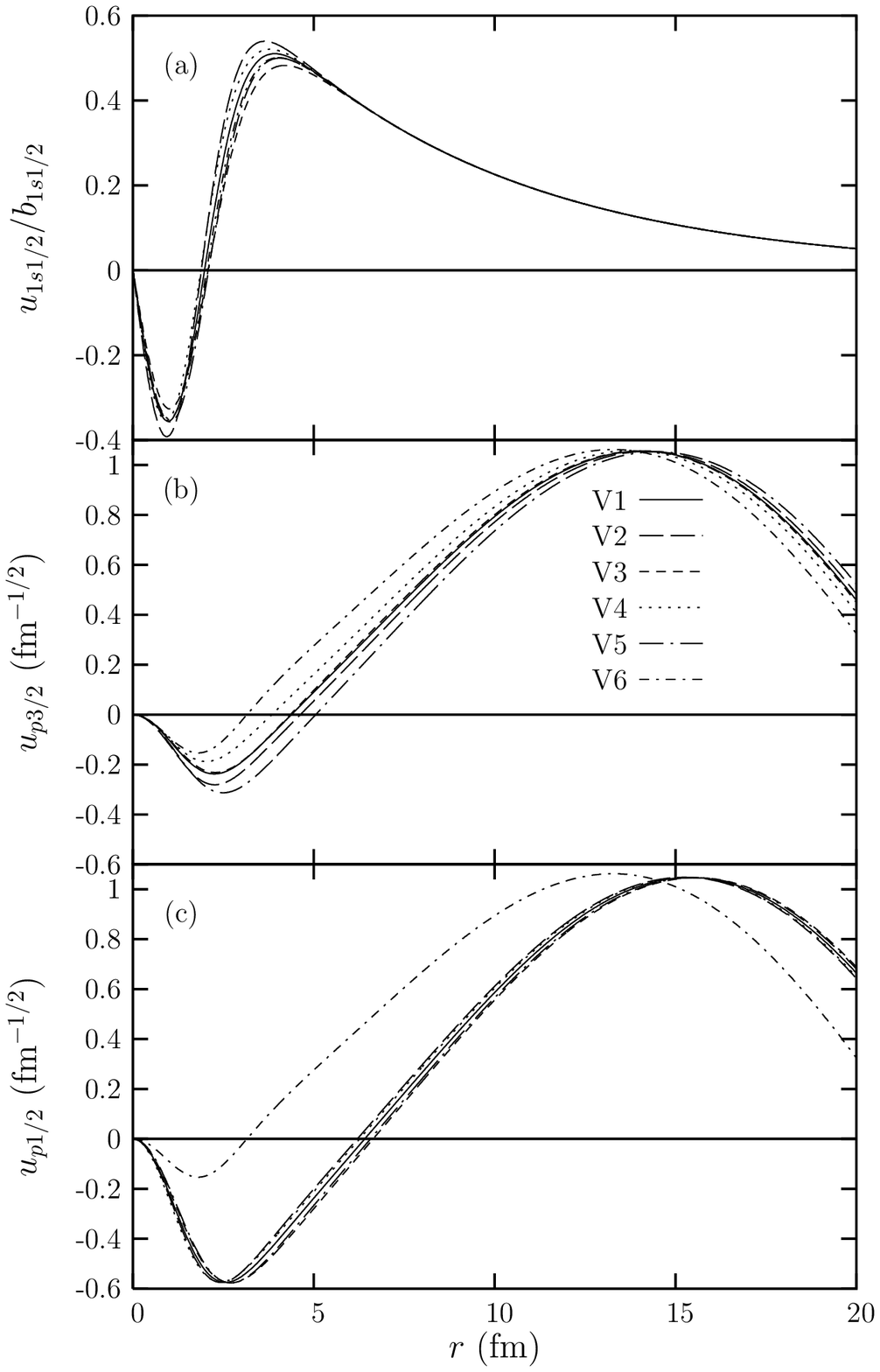}
\caption{Wave functions of \ex{11}Be obtained with the different
\ex{10}Be-$n$ potentials V1--V6:
(a) the initial ground state divided by $b_{1s1/2}$,
(b) and (c) the $p3/2$ and $p1/2$, respectively, continuum wave functions
computed at $E=1$~MeV.
}\label{f6}
\end{figure}

\fig{f6} depicts the radial wave functions of the initial
$1s1/2$ ground state (a), and of the $p3/2$ (b) and $p1/2$ (c)
scattering states at $E=1$~MeV.
They are calculated for the potentials V1--V6 of \tbl{t1}.
As in \fig{f5}(a), the ground state wave function is divided
by the ANC. 
As in the \ex{8}B case, due to the long range of the wave function,
and to the $r$ factor appearing in the matrix element,
the first order breakup cross section is not sensitive to
the variations at small radii.
The cross section should therefore be proportional to $b_{1s1/2}^2$.
However, as for \ex{8}B, the change in the potential leads
also to variations in the scattering wave functions.
In the present case, though, the changes in the dominant component
are much more significant than in the previous example, explaining the large differences
observed in the breakup cross section in \Sec{be11bu}.
We indeed observe significant variations
in the $p3/2$ phase shift [\fig{f6}(b)].
These variations are responsible for the large differences
in the $p3/2$ contribution to the breakup cross section
observed in \fig{f4}. They overcome by far the differences due
to the ANC, which means that the $^{11}$Be  breakup reaction of \Ref{Fuk04}
seems more sensitive to the way its continuum is described than to
the details of its initial single particle state.

The differences observed in the $p1/2$ contribution to the
cross section (\fig{f4}) can easily be understood by the variations
in the phase shift shown here.
The very similar $p1/2$ phase shifts obtained with
potentials V1--V5 justify that these five potentials
lead to approximately the same $p1/2$ contributions to
$(d\sigma/d\Omega)/{b_{1s1/2}^2}$.
The similarity in the $p1/2$ phase shifts
is due to the fact that all those potentials have been
adjusted in order to reproduce the very loosely bound
$\half^-$ excited state of \ex{11}Be in that partial wave.
Potential V6, that does not reproduce this excited state,
presents a very different phase shift of the $p1/2$ component, which
explains its larger $p1/2$ contribution to the cross section.

This first-order analysis of the breakup reactions
explains qualitatively the sensitivity of our calculations
to the core-fragment potential.
It shows that the wave functions describing both the ground state
and the continuum have a significant influence on the
cross section. Since the reaction is mostly peripheral,
this influence occurs mainly through the asymptotic
characteristics of the wave functions, i.e. the ANC
of the initial bound state, and the phase-shift in the
continuum.

These results confirm the study of the electromagnetic
strengths in one-nucleon halo nuclei
performed by Typel and Baur in \Ref{TB05}.
In that paper, the authors analyze the sensitivity
of $B(\rm{E}\lambda)$ transitions to the two-body
description of the loosely-bound nuclei. They show
that the energy distributions depend 
not only on the description of the bound state,
but also on that of the continuum.
Assuming the actual wave functions can be replaced by
their asymptotic behaviours, they obtain simple
analytical expressions for the energy distributions.
These expressions show clearly that the transition
strengths can be characterized by only a few parameters
describing the nucleus:
the binding energy of the system, the ANC of the
bound state wave function, and the scattering
length describing the final state interaction.
The present analysis shows that these conclusions
remain qualitatively valid in breakup reactions
where higher-order effects and nuclear interactions
are significant \cite{TNT01,CB05}.

It should be noted that the shape of the transition strengths
obtained
by Typel and Baur depends significantly on the scattering
length (see Figs.~9 and 10 of \Ref{TB05}). If this were so, it
should enable the extraction of the ANC and the scattering length
from the same data. However, our results show hardly any distortion of the
breakup cross sections with the potential choice.
The curves differ only in amplitude, not in shape, even though
the various potentials lead to significantly different phase shifts.
The consequence is then that the final state interaction
needs to be constrained by other observables in addition to
the breakup energy distribution.
This discrepancy between our results and those of \Ref{TB05}
is due to further approximations performed
in \Ref{TB05} required for obtaining analytic solutions. We find that
the analytic expressions of \Ref{TB05} are only useful
qualitatively. The shape and peak of the energy distributions
are not correctly produced.

\section{Conclusion}\label{conclusion}
In this paper, the breakup calculations of
two loosely bound projectiles, \ex{8}B and \ex{11}Be,
have been performed within CDCC and time-dependent frameworks
at low and high energy, respectively.
Different potentials were used to simulate the interaction between
both constituents of the projectile.
It has been shown that the potential geometry
has a significant influence upon the breakup cross section.
This dependence occurs through the asymptotic
normalization coefficient of the initial bound state,
as expected from the peripheral nature of the reaction.
However, the way excited states and the continuum are modeled
also plays a role. 

These results have been confirmed using DWBA and/or 
first-order perturbation theory.
These approximations enabled us to interpret the
variations observed in the more elaborated techniques with one-step theory.
Our analysis shows with transparency that different
potential geometries lead to differences
in the phase shifts that can affect significantly
the total cross section.
This qualitative analysis is in good agreement with
Typel and Baur's analytical study of the electromagnetic strengths
in halo nuclei \cite{TB05}.

One of our test cases consisted of the breakup of \ex{8}B 
on \ex{58}Ni at 25.75~MeV. 
Different potentials lead to minor variations in the
dominant $p3/2$ phase shift,
since all potentials have been adjusted to reproduce the \ex{8}B ground state
in that partial wave.
These variations induce only 4~\%  differences
in the breakup cross section, which remains therefore
approximately proportional to the square of the ANC.
On the contrary, our potentials  give large differences in the $p1/2$
phase shift, leading to significant distortion in the
energy distribution for this partial wave. As
this partial wave is less important in this reaction,
the phase shift differences have a smaller effect 
in the normalization of the total cross section.

In the breakup of \ex{11}Be on \ex{208}Pb at 69~MeV/nucleon,
no fitting was imposed on the dominant $p3/2$ partial wave, thus
the various potentials lead to significant
differences in the corresponding phase shifts.
These differences are so large that they overcome those
in the single particle ANC. The cross section
is therefore not proportional to the square of the ANC.
Five out of the six potentials we have used for this analysis
have been adjusted to reproduce the $\half^-$ excited state
of \ex{11}Be in the $p1/2$ partial wave. This induces
very similar phase shifts in that partial wave.
That contribution to the cross section is therefore nearly
proportional to the square of the ANC. A sixth potential,
that does not reproduce the excited state, has a very different
$p1/2$ phase shift, and leads to a much higher $p1/2$ contribution.

These results show that Coulomb breakup reactions probe not only
the ground state of the projectile, as it is usually assumed,
but also the continuum description.
We have indeed seen that differences in the phase shift
of the dominant partial wave
can influence significantly breakup calculations.
Therefore, one should be particularly cautious when 
extracting ground-state spectroscopic information or ANCs from
breakup measurements.
As shown in the calculation of the
Coulomb breakup of \ex{11}Be, different descriptions of
the continuum can lead to variations in the cross section
up to 40~\%, generating a significant inaccuracy in the
spectroscopic factor.
Moreover, even though the reaction is very peripheral,
the extraction of the single-particle ANC from breakup
measurements can be very tricky due to
the strong dependence of the calculations on the scattering waves.
The sensitivity of the extracted values to the potential geometry
should therefore be evaluated, and taken into account in the analysis.
Whenever possible,
the potentials used to simulate two-body projectiles should
be adjusted on other experimental data (excited states and phase shifts).
This underlines the need for additional data to constrain
the scattering properties of the core-fragment potentials.

\begin{acknowledgments}
We would like to thank A.~M.~Mukhamedzhanov for useful comments 
to the manuscript, and D.~Baye for interesting
discussions on this work.
P.~C. acknowledges the support of the Natural Sciences 
and Engineering Research Council of Canada (NSERC). Support
from the NSCL at Michigan State University and from the
National Science Foundation grant PHY-0456656 is
acknowledged.
\end{acknowledgments}


\end{document}